\documentclass[aip,reprint]{revtex4-2}

\usepackage{graphicx,siunitx,xcolor,hyperref,braket,bm,derivative,physics2}
\usepackage[version=4]{mhchem}
\usephysicsmodule{ab}
\hypersetup{
colorlinks,
linkcolor={blue!50!black},
citecolor={blue!50!black},
urlcolor={blue!80!black}}

\begin{document}
\title{Dynamic heterogeneity in sodium silicate melts via machine-learning potential}
\author{Kumpei Shiraishi}
\email{kumpei.shiraishi@sanken.osaka-u.ac.jp}
\author{Rikuta Nozawa}
\author{Emi Minamitani}
\affiliation{SANKEN, University of Osaka, 8-1 Mihogaoka, Ibaraki, Osaka 567-0047, Japan}
\date{\today}
\begin{abstract}
We present a comprehensive characterisation of dynamic heterogeneity in sodium silicate melts using molecular dynamics simulation with machine-learning potentials.
By studying sodium disilicate, tetrasilicate, and hexasilicate melts across a range of temperatures, mean squared displacement and a time-correlation function computed up to the nanosecond timescale provide a detailed account of how spatial mobility disparities emerge in a realistic multicomponent oxide glass.
Within these timescales, the self-part of the van Hove function for sodium displays a bimodality, demonstrating that alkali transport is mediated by discrete displacement events consistent with a hopping mechanism.
This distinct hopping allows sodium ions to decouple from the sluggish relaxation of the silicate matrix.
Furthermore, evaluation of the non-Gaussian parameter reveals that, although all constituent species exhibit dynamic heterogeneity, the non-Gaussian behaviour is most pronounced for oxygen atoms.
This trend reflects the intermittency of structural rearrangements, where framework atoms undergo rare and stochastic events compared to the frequent displacements of mobile ions.
Our findings elucidate the microscopic mechanism of ion transport and its connection to dynamic heterogeneity in silicate melts, offering a new avenue to study fundamental glassy physics in realistic vitreous materials.
\end{abstract}
\maketitle

\section{Introduction}
Ion diffusion in alkali-doped silica glasses is a fundamental problem of broad relevance, spanning chemical physics, materials science, and geology~\cite{Angell2000,Greaves_2007,10.2138/rmg.2022.87.04}.
Its importance has been further highlighted by applications in advanced industrial products such as all-solid-state secondary batteries, making it a timely and active area of research~\cite{10.3389/fenrg.2020.00218}, where precise control of ion transport is paramount.
The mechanism by which alkali ions relax within the disordered silicate network is a key factor governing the macroscopic properties of these materials.
At the heart of this context is the preferential pathway picture, which posits that alkali ions do not diffuse uniformly through the \ce{SiO2} matrix but instead migrate through spatially distinct, interconnected channels that form within the network~\cite{Greaves1985}.
To date, numerous studies have focused on verifying the static structural aspects of this framework, observing structural changes that support this depiction as sodium density increases~\cite{PhysRevB.52.6358,Meyer_2002,Meyer2004Channel}.
However, since ion transport is inherently a dynamical phenomenon, there is a pressing need to characterise these processes from the perspective of relaxation dynamics through direct observation via computer simulations~\cite{HORBACH200187,Jund2001,PhysRevLett.88.125502,Sorensen2023}.

Such a characterisation is most naturally pursued within the broader framework of glass physics, where relaxation dynamics have long been a central concern~\cite{Ediger_1996,Debenedetti_2001,Berthier_Biroli_2011}.
In supercooled liquids approaching the glass transition, particles of high and low mobility coexist and fluctuate spatially, a phenomenon known as dynamic heterogeneity~\cite{Ediger_2000}.
These dynamical fluctuations grow as the system cools and relaxation slows, giving rise to hallmark features of the glass transition such as stretched exponential decay of time correlation functions~\cite{Kob_Andersen_PRL_1994,Kob_Andersen_I_1995,Kob_Andersen_II_1995}, exponential distribution tails of single-particle displacements~\cite{Chaudhuri_2007}, and four-point susceptibility peaks~\cite{Lacevic2003,Berthier_2005,Toninelli_2005}.
Moreover, dynamic heterogeneity is not merely of theoretical interest;
it governs transport properties, including macroscopic diffusion~\cite{Cicerone_1996,Sengupta2013JCP,Kawasaki_2017} and ionic conductivity~\cite{Horbach1999PRB,Habasaki2007MAE}, and is therefore of direct relevance to the ion migration picture outlined above.

Nevertheless, detailed insights into dynamic heterogeneity have been predominantly obtained from packing-type glass models, in which interparticle interactions are isotropic and short-ranged~\cite{Kob_Andersen_I_1995,Kob_DH_1997,Donati_1998,Weeks_2000}.
The correlation between this phenomenon and ion diffusion in network glasses such as sodium silicate, where directional covalent bonding gives rise to qualitatively richer local environments, remains to be fully elucidated.
It is therefore essential to accurately capture individual particle dynamics within the diverse local environments that are far more pronounced in network glasses, typified by sodium silicate, than in simple packing-type systems.
From this perspective, it is crucial to go beyond simple empirical potentials~\cite{HORBACH200187,PhysRevLett.88.125502,Meyer2004Channel} and track relaxation with the precision of first-principles calculations that incorporate the underlying electronic states.
However, as is well known, first-principles calculations are computationally intensive, making it increasingly impractical to cover the growing relaxation times near the glass transition as the temperature decreases.
More critically, the computational cost of first-principles methods scales as $O(N^3)$ with respect to the number of particles.
This scaling is fundamentally at odds with the requirement for large system sizes necessary to capture dynamic heterogeneity in glass transition research.

In this study, we overcome this computational barrier by utilising machine learning potentials, which have undergone remarkable methodological advances~\cite{Behler_2007,Deringer2019}.
Such potentials have also enabled detailed and transferable atomistic modelling of the silicon-oxygen system, spanning crystalline, amorphous, and nanostructured environments~\cite{Erhard2022,Erhard2024}, including alkali silicate glasses closely related to the compositions studied here~\cite{Bertani2024,Pedone2025,Ganisetti2025}.
Whereas these prior works on alkali silicate glasses have focused primarily on establishing the accuracy and transferability of machine-learning force fields through structural validation, the present study extends this approach to the long-time dynamical regime.
Specifically, through the construction of a machine learning potential for sodium silicate and a molecular dynamics (MD) code developed in-house and optimised for graphics processing unit execution, we have succeeded in dynamically tracking the relaxation of both the frozen silica network and the diffusive sodium ions over extended spatial and temporal scales previously inaccessible to first-principles methods.
This computational strategy, which leverages the architectural characteristics of modern hardware, enables the application of analytical methods developed in fundamental glass transition research to realistic sodium silicate systems.
We provide compelling evidence for the preferential pathway picture from a dynamical perspective, supported by complementary structural evidence presented in the Appendix, establishing a new foundation for the study of realistic glass systems.

\section{Methods}
The machine-learning interatomic potential (MLIP) was constructed using a PaiNN-type equivariant graph neural network~\cite{Schutt2021-py}.
Atomic features were embedded in a 60-dimensional latent space, and interatomic interactions were evaluated within a cutoff radius of \qty{5.0}{\angstrom} using 40 Gaussian radial basis functions combined with a smoothstep cutoff envelope.
The network architecture comprised two interaction blocks.

The model parameters were optimised by minimising a weighted loss function incorporating total energies, atomic forces, and cell virials:
\begin{align}
\mathcal{L} = w_E \mathcal{L}_E + w_F \mathcal{L}_F + w_{\Xi} \mathcal{L}_{\Xi},
\end{align}
where the weighting factors were set to $w_E = 0.01$, $w_F = 0.99$, and $w_{\Xi} = 0.01$.
The energy, force, and cell-virial contributions were evaluated as mean squared errors relative to the reference first-principles values.
To ensure high-fidelity MD trajectories, a substantially larger weight was assigned to the force-loss term.

The training dataset was generated from ab initio molecular dynamics melt-quench trajectories of sodium silicate glasses.
Initial configurations were prepared for four \ce{Na2O} fractions, corresponding to \ce{Na6Si57O117}, \ce{Na12Si54O114}, \ce{Na24Si48O108}, and \ce{Na48Si36O96} in 180-atom cells, each prepared at four mass densities of \qtylist{1.8; 2.0; 2.2; 2.4}{\gram\per\cm\cubed}, yielding 16 density-composition conditions in total.

The ab initio molecular dynamics simulations were performed using VASP~\cite{Kresse1993,Kresse1996CMS,Kresse1996PRB} with the PBE functional~\cite{Perdew1996} and PAW potentials~\cite{Kresse1999}.
The plane-wave cutoff energy was set to \qty{400}{\electronvolt}, and only the $\Gamma$-point was sampled in $k$-space.
Each system was first run at \qty{5000}{\kelvin} for \qty{3}{\pico\second} and subsequently quenched to \qty{300}{\kelvin} over \qty{4}{\pico\second} using a \qty{1}{\femto\second} time step.

Atomic configurations were sampled every 10 MD steps from both the high-temperature equilibration and the subsequent quenching trajectories, yielding \num{699} configurations for each density--composition condition and a total dataset of \num{11184} structures.
For each sampled configuration, the total energy, atomic forces, and cell virials were recomputed through static single-point DFT calculations with accurate precision settings.

The resulting dataset was randomly partitioned into training (\qty{80}{\percent}) and validation (\qty{20}{\percent}) sets.
The MLIP was trained for 500 epochs using the AdamW optimiser~\cite{Loshchilov2018-hh} with an initial learning rate of $10^{-3}$, which was reduced by a factor of 0.5 every 100 epochs using a step scheduler.
The validation accuracy of the trained potential is presented in Appendix~\ref{sec:model_accuracy}.

Production runs with the trained MLIP were carried out for sodium silicate systems containing \num{3000} atoms.
The linear size of the simulation box was determined by a mass density of \qty{2.30}{\gram\per\cm\cubed}.
Three distinct compositions were investigated: \ce{Na666Si667O1667}, \ce{Na400Si800O1800}, and \ce{Na284Si858O1858}, which approximately correspond to \ce{Na2O} $\cdot$ \ce{2SiO2}, \ce{Na2O} $\cdot$ \ce{4SiO2}, and \ce{Na2O} $\cdot$ \ce{6SiO2} (hereafter designated as NS2, NS4, and NS6, respectively).
For each composition, three independent initial configurations were generated via random packing.

The systems were initially equilibrated in the liquid state at \qty{4000}{\kelvin} for \qty{20}{\pico\second} using the Bussi thermostat~\cite{Bussi_2007}, and subsequently cooled to \qty{300}{\kelvin} at a constant cooling rate of \qty{e12}{\kelvin\per\second}.
During this cooling process, instantaneous configurations were sampled at target temperatures of \qtylist{4000; 3400; 2900; 2500; 2100; 1900; 1700; 1500}{\kelvin}.
At each temperature, an additional NVT run of \qty{20}{\pico\second} was performed using the Nosé--Hoover thermostat~\cite{Nose_JCP_1984,Hoover1985}, followed by production runs in the NVT ensemble.
The durations of production runs were:
\qty{5.1}{\nano\second} for \qty{1500}{\kelvin},
\qty{3.2}{\nano\second} for \qty{1700}{\kelvin},
\qty{1.3}{\nano\second} for \qty{1900}{\kelvin} and \qty{2100}{\kelvin},
\qty{0.51}{\nano\second} for \qty{2500}{\kelvin},
\qty{0.32}{\nano\second} for \qty{2900}{\kelvin}, and
\qty{0.13}{\nano\second} for \qty{3400}{\kelvin} and \qty{4000}{\kelvin}.
These extended production windows were selected to cover the relevant structural relaxation timescales, particularly for the network-forming species at lower temperatures.

\section{Bulk relaxations}
\label{sec:relaxation}

\subsection{Mean squared displacement}
In this section, we discuss the bulk relaxation of the system and its characteristic relaxation time.
We first examine the mean squared displacement (MSD), which is defined as follows:
\begin{align}
\braket{r^2(t)} = \Braket{\frac{1}{N_\alpha} \sum_{i=1}^{N_\alpha} \lvert \bm{r}_i(t) - \bm{r}_i(0) \rvert^2},
\end{align}
where $N_\alpha$ represents the number of $\alpha$-atoms in the system.
MSD was evaluated for each atomic species in the system for every composite and temperature.

\begin{figure*}
\centering
\includegraphics[width=\linewidth]{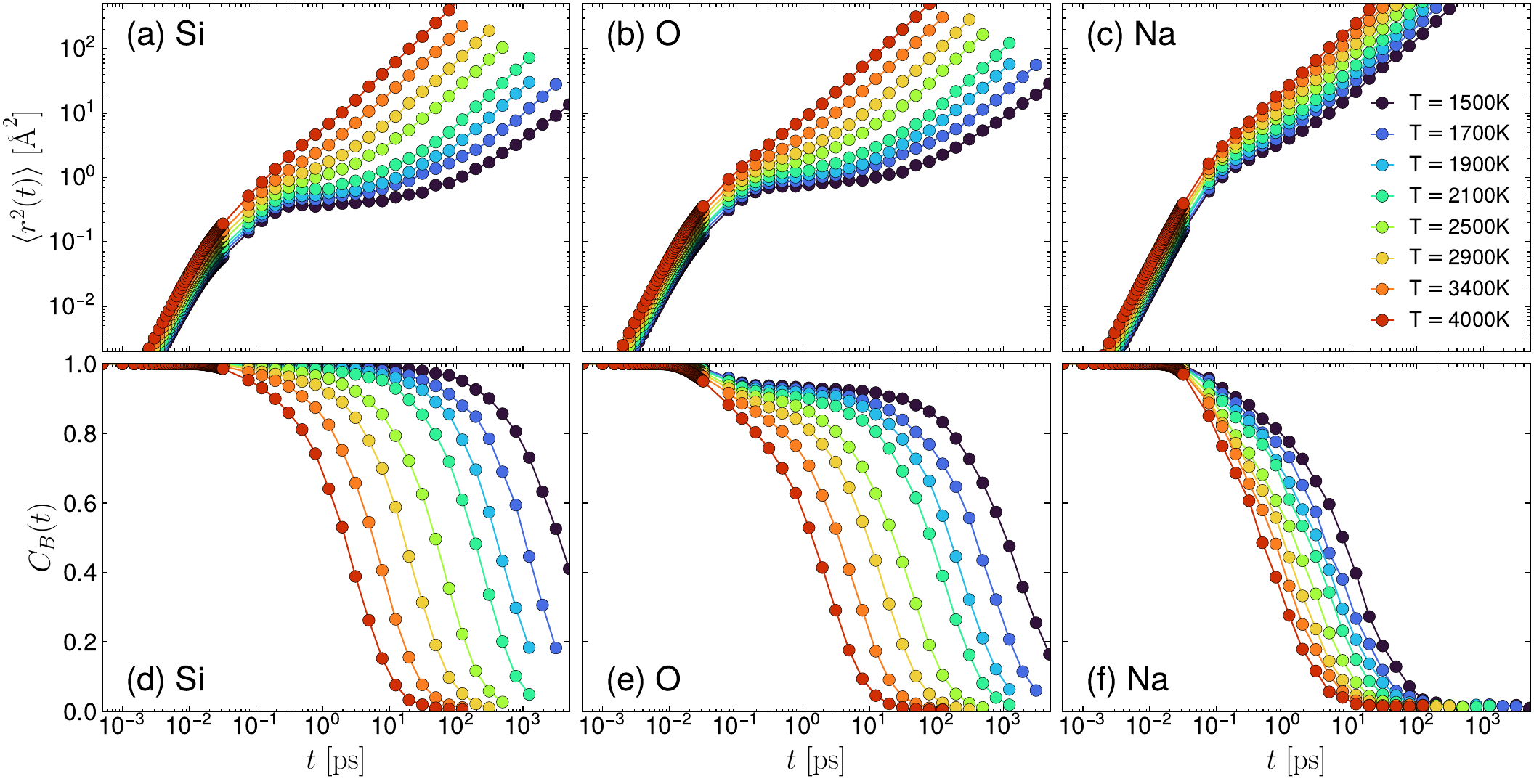}
\caption{Mean squared displacements $\braket{r^2(t)}$ (panels (a--c)) and bond-break correlation $C_B(t)$ (panels (d--f))of NS4 composite.
Panels (a, d), (b, e), and (c, f) represent data from \ce{Si}, \ce{O}, and \ce{Na} atoms, respectively.}
\label{fig:MSD}
\end{figure*}

As illustrated in Fig.~\ref{fig:MSD} (a--c) for the NS4 composition, at the highest temperature $T = \qty{4000}{\kelvin}$, a smooth transition from the short-time ballistic behaviour to long-time diffusive behaviour is observed for all atomic species (\ce{Si}, \ce{O}, and \ce{Na}).
Upon cooling, the MSD curves exhibit characteristic changes.
For the \ce{Si} and \ce{O} atoms composing the \ce{SiO2} framework, a distinct plateau emerges at intermediate timescales.
This feature of MSD represents the two-step relaxation process, a hallmark of the glass transition phenomenon~\cite{Kob_Andersen_I_1995}.
Even at low temperatures where this plateau is pronounced, our time window covers the diffusive regime of MSD of \ce{Si} and \ce{O} atoms at sufficiently long times.
In contrast, the relaxation of \ce{Na} atoms is considerably more rapid than that of the species of \ce{SiO2} framework.
Notably, even at the lowest temperature investigated, $T = \qty{1500}{\kelvin}$, an obvious plateau does not emerge in the MSD of \ce{Na}, unlike the cases of \ce{Si} or \ce{O}.
Nevertheless, as the temperature decreases, its MSD begins to exhibit incipient signs of a plateau in the intermediate time regime $t \approx \qty{1}{\pico\second}$, reflecting the onset of slow dynamics of this species at lower temperatures.
Taken together, these results demonstrate that our simulations succeed in characterising the relaxation dynamics of a complex oxide melt over nanosecond timescales.
This temporal regime remains entirely inaccessible to ab initio molecular dynamics and has only recently begun to be reached by machine-learning potentials for simpler supercooled liquids, such as toluene~\cite{Pabst2025}.

\subsection{Bond-break correlation function}
Next, we present the results for the bond-break correlation function~\cite{Yamamoto_1997,Yamamoto_1998}
\begin{align}
C_B(t) = \Braket{ \frac{1}{N_\alpha} \sum_{i=1}^{N_\alpha} \frac{N_B^i(t \vert 0)}{N_B^i(0)} },
\end{align}
where $N_B^i(0)$ represents the number of neighbouring $\alpha$-atoms at time $t = 0$ and $N_B^i(t \vert 0)$ represents the number of the neighbouring $\alpha$-atoms that remain bonded at time $t$.
The neighbouring two particles $i$ and $j$ are determined by whether their distance $r_{ij}$ is smaller than a threshold value of their species $\alpha$: $r_{ij} < r_{\mathrm{th}, \alpha}$.
The thresholds $r_\alpha$ are selected as the first minimum of the pair correlation function (shown in Supplementary Information), namely $r_\mathrm{th, \ce{Si}} = 3.6$, $r_\mathrm{th, \ce{O}} = 3.2$, and $r_\mathrm{th, \ce{Na}} = 5.0$.
The threshold at time $t$ is set to $1.1 r_{\mathrm{th}, \alpha}$.
This function represents the probability that pairs of atoms, initially located within each other's proximity, remain within a specified threshold distance at time $t$.
This correlation function is shown to serve as a good time-correlation function for capturing relaxations of supercooled liquids~\cite{Shiba_2012}.

$C_B(t)$ is shown in Figs.~\ref{fig:MSD} (d--f) for the NS4 composition in correspondence with the MSD.
The temporal evolution of $C_B(t)$ exhibits a trend consistent with that of the MSD.
At high temperatures, a single-step relaxation is evident, where the correlations for all atomic species (\ce{Si}, \ce{O}, and \ce{Na}) vanish within approximately \qty{10}{\pico\second}.
For \ce{Si} and \ce{O}, the substantial slowing down of relaxation at lower temperatures is clearly captured by $C_B(t)$.
In the low-temperature regime, a well-defined plateau develops for these species, indicative of the two-step relaxation process that is a typical characteristic of the glass transition~\cite{Kob_Andersen_I_1995}.
At short times, $C_B(t)$ undergoes an initial decay associated with the $\beta$ relaxation process before entering the plateau region.
This first decay to the plateau is notably more pronounced for \ce{O} than for \ce{Si}.
The higher clarity of the first relaxation step and the plateau for oxygens can be attributed to the specific local environment of the species;
as oxygen atoms constitute the vertices of the \ce{SiO4} tetrahedra, they are more directly susceptible to the local cage effect formed by neighbouring atoms of the same type.
Consequently, the short-time $\beta$ relaxation, which is considered as thermal vibrations inside cages~\cite{Kob_Andersen_I_1995}, and caging by surrounding atoms are more readily observable for \ce{O} atoms than for central \ce{Si} atoms.
On the other hand, at sufficiently long times, the bond-break correlation eventually decays to zero ($\alpha$ relaxation), which corresponds to the transition to the diffusive regime observed in the MSD.

In contrast, the slowdown in the relaxation of \ce{Na} atoms is relatively modest, even as the temperature decreases to the regime where relaxations of \ce{Si} and \ce{O} atoms are already slow.
At the lowest temperature investigated ($T = \qty{1500}{\kelvin}$), while the \ce{Si} and \ce{O} correlations persist for roughly \qty{1}{\nano\second} before decaying, the \ce{Na} correlation is lost within \qtyrange{10}{100}{\pico\second}, approximately one to two orders of magnitude faster than the silica framework.
Nevertheless, a discernible slowdown in the \ce{Na} dynamics is still observed at low temperatures (e.g.~$T = \qty{1500}{\kelvin}$), with incipient signs of a plateau appearing at $t \approx \qty{1}{\pico\second}$.

\subsection{Structural relaxation time}
\begin{figure}
\centering
\includegraphics[width=\linewidth]{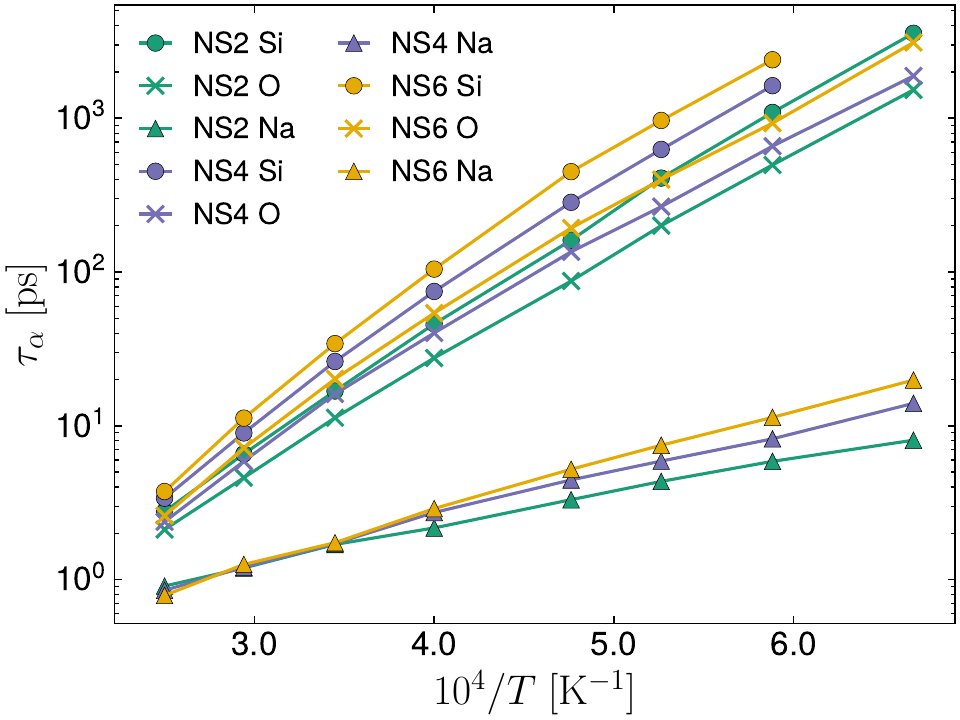}
\caption{Relaxation time $\tau_\alpha$ of all compositions.
For \ce{Si} atoms at the lowest temperature, $C_B(t)$ does not decay to the threshold value $e^{-1}$ within the simulation window (see Fig.~\ref{fig:MSD} (d) for the NS4 case);
these data points are therefore absent.}
\label{fig:tau_alpha}
\end{figure}

Utilising the bond-break correlation functions, we now define the structural relaxation time $\tau_\alpha$ as $C_B(\tau_\alpha) = e^{-1}$.
Figure~\ref{fig:tau_alpha} presents the temperature-dependence of these relaxation times for each constituent species.
Across all investigated compositions, the increase of relaxation times of the network-forming \ce{Si} and \ce{O} atoms is manifestly more sluggish than that of the mobile \ce{Na} ions.
The disparate sensitivity of these relaxation times to temperature, evident in the steeper gradients of $\tau_\alpha$ for \ce{Si} and \ce{O}, indicates that the dynamic decoupling between the \ce{SiO2} framework and the alkali ions becomes increasingly pronounced as the temperature decreases.
Within the network itself, a clear hierarchy is observed, with \ce{Si} atoms consistently exhibiting longer relaxation times than \ce{O} atoms.
Notably, the reliable extraction of $\tau_\alpha$ for \ce{Si} and \ce{O} down to $T = \qty{1500}{\kelvin}$ was made possible only by the extended simulation timescale afforded by the present machine-learning potential, as such temperatures correspond to relaxation times beyond the reach of earlier first-principles studies.

When comparing the dynamics across the various compositions, it is observed that an increase in sodium leads to a systematic acceleration of the dynamics for all atomic species.
This compositional influence on the relaxation behaviour is most salient in the low-temperature regime.
Conversely, at high temperatures, the disparity between the compositions diminishes;
notably, the relaxation times for \ce{Na} converge, displaying a near-total independence from the specific composition.
In contrast, the dynamics of \ce{Si} and \ce{O} remain sensitive to the network environment even in the high-temperature limit, maintaining a degree of compositional dependence that is absent in the alkali ion.

\section{Dynamic heterogeneity}
\label{sec:DH}
\subsection{The van Hove function}
\begin{figure}
\centering
\includegraphics[width=\linewidth]{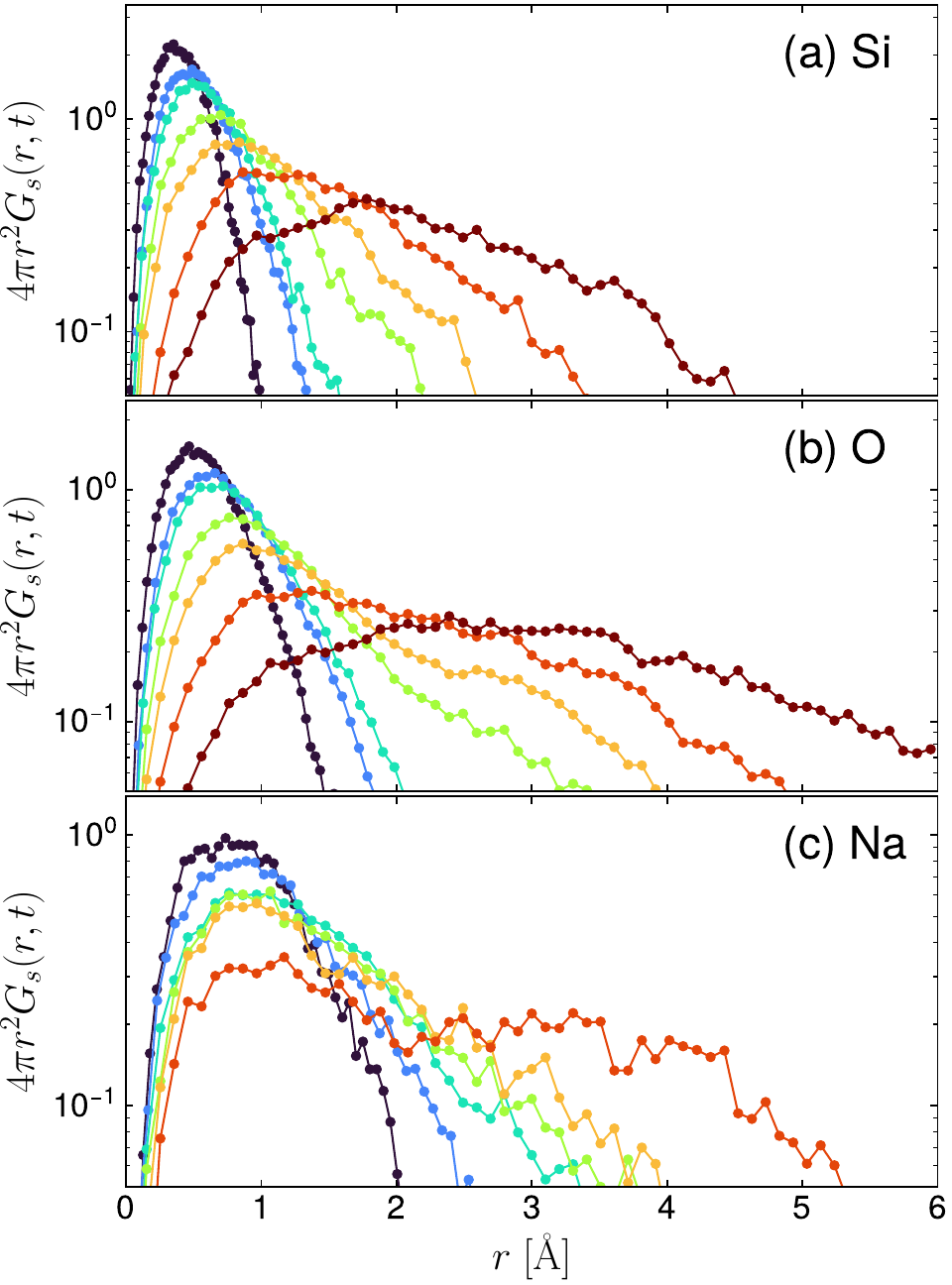}
\caption{Self-part of the van Hove distribution function for the NS4 composite at \qty{1500}{K}.
Panels (a), (b), and (c) represent data for \ce{Si}, \ce{O}, and \ce{Na} species, respectively.
Each data corresponds to \qtylist{0.122;1.230;12.299;122.995;308.949;776.046;1949.339}{\pico\second} for \ce{Si} and \ce{O} atoms and \qtylist{0.122;0.195;0.489;0.775;1.230;4.896}{\pico\second} for \ce{Na} atoms.}
\label{fig:vanHove}
\end{figure}

To further elucidate the nature of the relaxation processes, we now turn our attention to the spatial heterogeneity of the relaxation.
We first examine the spatial distribution of atomic displacements via the self-part of the van Hove distribution function~\cite{HansenMcDonald}, defined as:
\begin{align}
G_s^\alpha(r,t) = \Braket{\frac{1}{N_\alpha} \sum_{i=1}^{N_\alpha} \delta\pab{r - \lvert \bm{r}_i(t) - \bm{r}_i(0) \rvert}},
\end{align}
where $\delta(x)$ denotes the Dirac delta function.
In the following discussion, the distribution is presented as $4\pi r^2 G_s^\alpha(r, t)$ to account for the three-dimensional volume element.

Figure~\ref{fig:vanHove} illustrates the temporal evolution of this distribution for the NS4 composite at \qty{1500}{K}.
At short times, the distributions show the Gaussian property for each atom.
On the other hand, at intermediate and long timescales, the network-forming species, namely \ce{Si} and \ce{O} atoms (panels (a) and (b)), exhibit the distributions with pronounced non-Gaussian tails.
These tails signify a substantial departure from the Gaussian behaviour typical of simple liquids, indicating that a subpopulation of these atoms undergoes displacements significantly larger than the typical distribution.

In contrast, the mobile \ce{Na} ions (panel (c)) display a markedly different dynamical profile.
At extended timescales, the distribution exhibits a discernible bimodality, a hallmark of discrete hopping dynamics, wherein the secondary peak corresponds to ions that have successfully escaped their local coordination cages and migrated to adjacent interstitial sites.
This peak is located at approximately $r \approx \qty{3.5}{\angstrom}$, consistent with the typical \ce{Na}--\ce{Na} nearest-neighbour distance derived from the partial pair correlation function (Supplementary Information), confirming that these displacement events correspond to discrete jumps between adjacent sites within the silicate network.
The emergence of such bimodality provides direct evidence for the coexistence of relatively immobile, caged ions and highly mobile, jumping ions, and is a key manifestation of dynamic heterogeneity in alkali silicate systems.

\subsection{Non-Gaussian parameter}
In the previous section, our analysis of the van Hove distribution revealed the non-Gaussian nature of the relaxation and the separation of mobile and immobile atoms within the sodium silicate.
To quantitatively characterise this heterogeneity, we next evaluate the non-Gaussian parameter $\alpha_2(t)$~\cite{Rahman1964}.
In three dimensions, this is defined as:
\begin{align}
\alpha_2 (t) = \frac{3\braket{r^4(t)}}{5\braket{r^2(t)}^2} - 1.
\end{align}
This parameter vanishes for a purely Gaussian process, and its magnitude serves as a probe of the deviation from the Gaussian distribution, thereby reflecting the spatial heterogeneity in particle mobilities.

\begin{figure*}
\centering
\includegraphics[width=\linewidth]{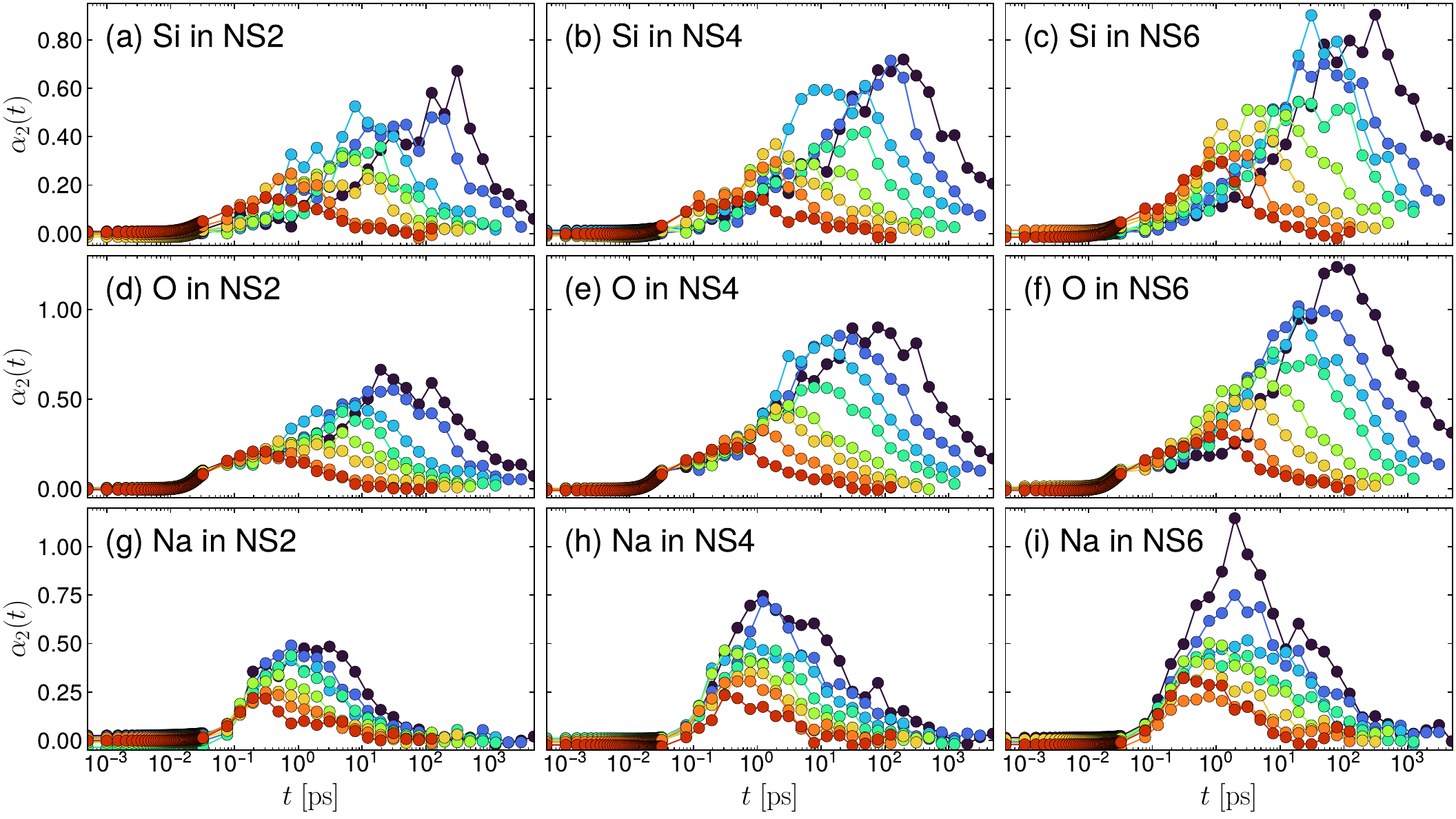}
\caption{Temperature and compositional dependence of the non-Gaussian parameter $\alpha_2(t)$ for every constituent species.
Panels (a--c), (d--f), and (g--i) represent data for \ce{Si}, \ce{O}, and \ce{Na} species, respectively.
Panels (a), (d), and (g) represent data for NS2, panels (b), (e), and (h) represent data for NS4, and panels (c), (f), and (i) represent data for NS6.
The legend of each panel is the same as that in Fig.~\ref{fig:MSD}.}
\label{fig:alpha2}
\end{figure*}

Figure~\ref{fig:alpha2} presents the evolution of $\alpha_2(t)$ for \ce{Si}, \ce{O}, and \ce{Na} across all investigated compositions (NS2, NS4, and NS6).
For all species and compositions, $\alpha_2(t)$ exhibits a characteristic peak at intermediate timescales within our time window.
The peak height increases systematically as the temperature is lowered, underscoring the intensification of dynamic heterogeneity as the system approaches the glass transition.
The position of this peak shifts towards longer timescales at lower temperatures, mirroring the overall slowing down of the structural relaxation.
We note, however, that the characteristic timescale of the $\alpha_2(t)$ peak remains consistently shorter than the structural relaxation time $\tau_\alpha$ defined from the time-correlation function, as observed in other common glass-forming models~\cite{Kob_Andersen_I_1995,Sciortino_1996,Donati_PRE_1999,PhysRevE.64.051503,Starr_2013,Das2022}.

A distinct compositional trend is also evident.
As the sodium content decreases from NS2 to NS6, corresponding to an increase in the connectivity of the silicate network (see Appendix~\ref{sec:static_structure}), the maximum value of $\alpha_2(t)$ increases for all atomic species.
This indicates that a reduction in modifier concentration enhances the spatial constraints imposed by the rigid framework, thereby amplifying the intermittency of relaxation events and separation of mobile and immobile particles.

Notably, the peak intensity of $\alpha_2(t)$ for oxygen is consistently the most pronounced among all constituent species across the investigated range.
In contrast, the magnitudes for silicon and sodium are comparable, with the sodium peaks often reaching or exceeding those of silicon.
The elevated non-Gaussianity in the oxygen dynamics stems from the structural and chemical diversity of its local environment.
While the central silicon atoms are rigidly anchored within the \ce{SiO4} units, the vertex oxygen atoms possess greater vibrational and rotational flexibility, allowing for more diverse local relaxation modes.
Furthermore, the coexistence of bridging and non-bridging oxygens in the presence of sodium modifiers introduces a stark contrast in local mobilities, thereby broadening the overall distribution of relaxation times for the oxygen species.

For sodium, the robust peak intensity is intrinsically linked to its specific transport mechanism.
As evidenced by the bimodality in the self-part of the van Hove functions, which is a feature absent in the framework atoms, sodium ions undergo discrete hopping events between interstitial sites.
The significant magnitude of $\alpha_2(t)$ for sodium thus reflects the highly intermittent nature of these hops, representing a sharp departure from continuous Gaussian diffusion.
Collectively, these results indicate that the dynamical heterogeneity in sodium silicate melts arises from two distinct origins:
the infrequent and stochastic rearrangements of the covalent network and the rapid, channel-mediated hopping of the alkali modifiers.

\section{Conclusions}
In this study, we have characterised the dynamical heterogeneity in sodium silicate melts across a wide range of temperatures and compositions through the application of machine-learning potentials with ab initio accuracy.
The primary significance of this work lies in our ability to extend high-precision simulations to timescales that fully encompass the structural relaxation of the silicate framework.
This has enabled a definitive account of how spatial heterogeneities in mobility manifest in a realistic multi-component oxide glass, bridging the gap between first-principles accuracy and the underlying physics behind the complex relaxations near the glass transition.
The structural origins of this behaviour are detailed in the Appendix~\ref{sec:static_structure}, where we demonstrate that the partial structure factors and $Q(n)$ species distributions are in good agreement with prior experimental and simulation studies of sodium silicate systems~\cite{Meyer2004Channel,HORBACH200187}.
In particular, the growth of the prepeak in $S_\mathrm{NaNa}(k)$ with increasing sodium content and the systematic shift of the $Q(n)$ distribution towards fragmented species provide direct structural evidence for the depolymerisation of the silicate network and the formation of sodium-rich preferential channels, corroborating the dynamical picture presented in the main text.

Our results provide a robust characterisation of dynamic heterogeneity through the systematic evaluation of the non-Gaussian parameter $\alpha_2(t)$.
We have demonstrated that the intensity and timescale of these dynamical fluctuations are highly species-dependent, with the oxygen atoms consistently exhibiting the most pronounced non-Gaussianity.
The magnitude of $\alpha_2(t)$ for sodium ions is found to be comparable to, and at certain temperatures exceeds, that of the central silicon atoms.
This observation reflects the distinct origins of heterogeneity in this system:
the rare, stochastic rearrangements of the rigid silicate framework and the discrete, intermittent hopping events of the alkali modifiers within the interstitial space.

The achievement of ab initio-level accuracy was instrumental in capturing the subtle interplay between the structural relaxation of the matrix and the motion of the modifiers.
We identified a bimodality in the self-part of the van Hove functions for sodium, providing evidence of hopping-mediated transport that persists even as the surrounding framework approaches a frozen state.
This decoupling, where the alkali dynamics remain non-Gaussian while the framework undergoes slow structural relaxation, represents a fundamental characteristic of the relaxations of modified silicate melts that can be accurately resolved through the high-accuracy molecular dynamics simulation.

In conclusion, our work demonstrates that the characterisation of dynamical heterogeneity in realistic glass-forming systems can be elevated to the level of near-ab initio accuracy.
By reaching the structural relaxation regime while maintaining the accuracy of the underlying atomic interactions, we have established a rigorous basis for understanding how local constraints and chemical environments dictate the macroscopic transport properties of multi-component oxide glasses.
These insights offer a new standard for the computational design and analysis of ion-conducting materials where the precise control of dynamical fluctuations is of paramount importance.

\begin{acknowledgments}
This work is supported by JSPS KAKENHI Grant Number 25H01478, JST FOREST Grant Number JPMJFR236Q, and a grant from the Inamori Foundation.
The computation was performed using computational facilities at the Research Center for Computational Science, Okazaki, Japan (Project: 26-IMS-C117).
\end{acknowledgments}

\section*{Conflict of Interest}
The authors have no conflicts to disclose.

\section*{Author contributions}
\textbf{K.~Shiraishi:} Conceptualization, Data curation, Formal analysis, Investigation, Methodology, Project administration, Software, Validation, Visualization, Writing - original draft, Writing - review \& editing;
\textbf{R.~Nozawa:} Software, Validation, Writing - review \& editing;
\textbf{E.~Minamitani:} Conceptualization, Data curation, Funding acquisition, Methodology, Project administration, Resources, Supervision, Validation, Visualization, Writing - review \& editing.

\appendix
\section{Model accuracy}
\label{sec:model_accuracy}
\begin{figure*}
\centering
\includegraphics[width=.325\linewidth]{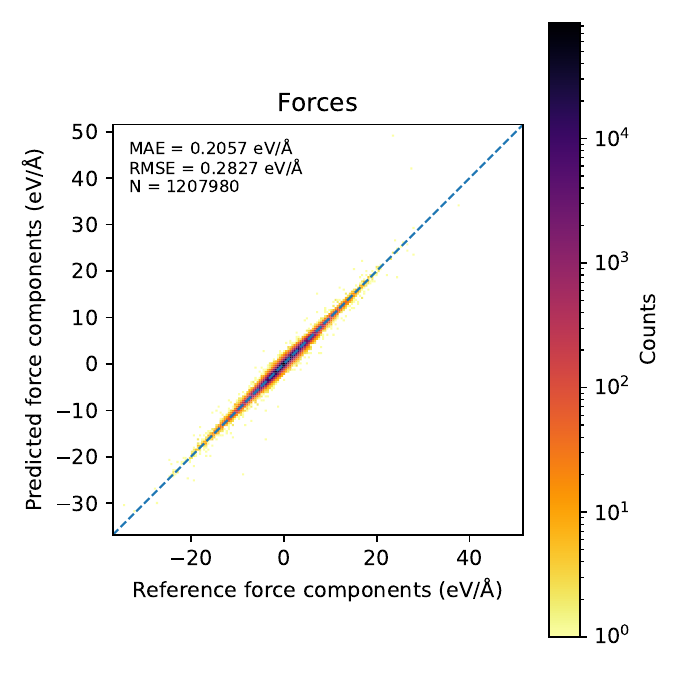}
\includegraphics[width=.325\linewidth]{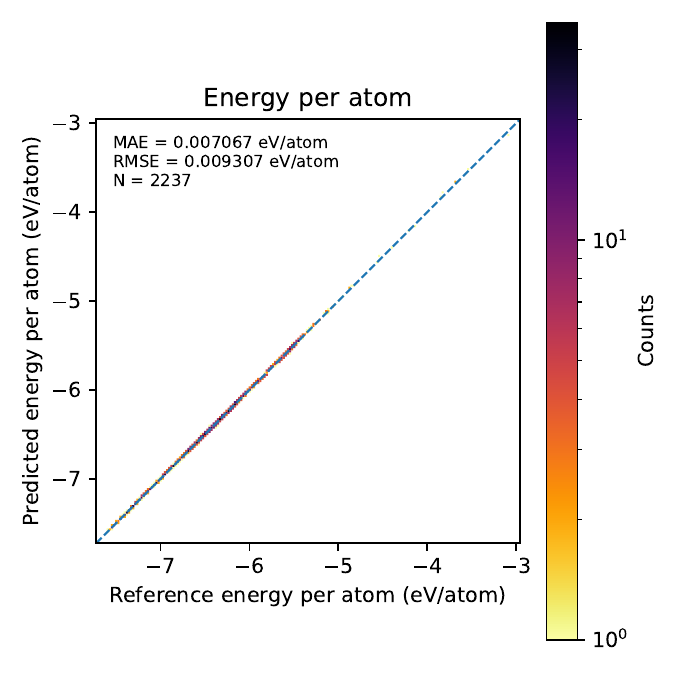}
\includegraphics[width=.325\linewidth]{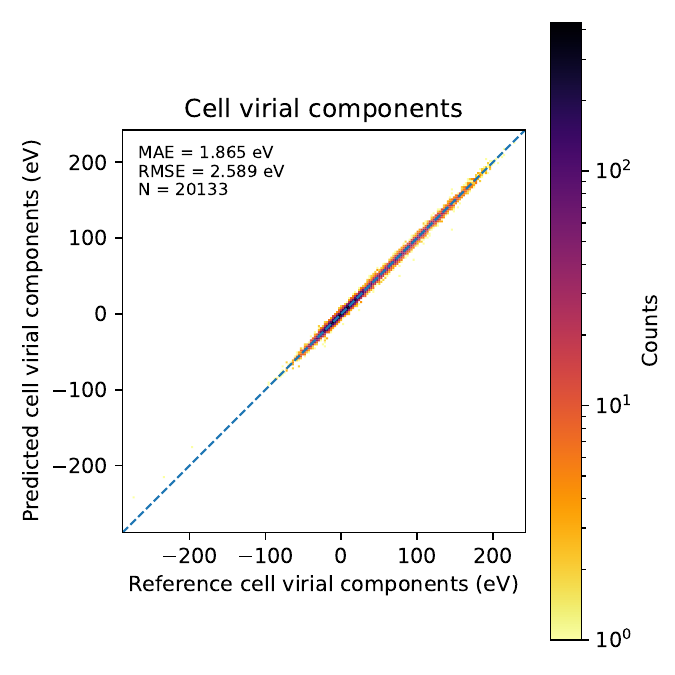}
\caption{Parity plot of predicted versus reference atomic forces, energies per atom, and cell virials (from left to right) on the held-out validation set.}
\label{fig:parity}
\end{figure*}

To assess the accuracy of the trained machine-learning interatomic potential, we compare the model predictions against the reference DFT values on the held-out validation set.
Figure~\ref{fig:parity} shows the parity plots for atomic forces, energies per atom, and cell virials.
The close agreement between predicted and reference values across the full range of the validation set demonstrates that the trained potential reproduces the underlying first-principles energy, force, and virial surfaces with good accuracy.

\section{Static structures}
\label{sec:static_structure}
In this appendix, we present a structural characterisation of the sodium silicate melts that complements the dynamical analysis presented in the main text.
We examine the partial structure factors $S_{\alpha\beta}(k)$ and the distribution of bond-bridging oxygen species $Q(n)$ across the investigated temperatures and compositions, providing the structural basis for the topological fragmentation and kinetic decoupling discussed in Secs.~\ref{sec:relaxation} and \ref{sec:DH}.

\subsection{Static structure factors}
\begin{figure}
\centering
\includegraphics[width=\linewidth]{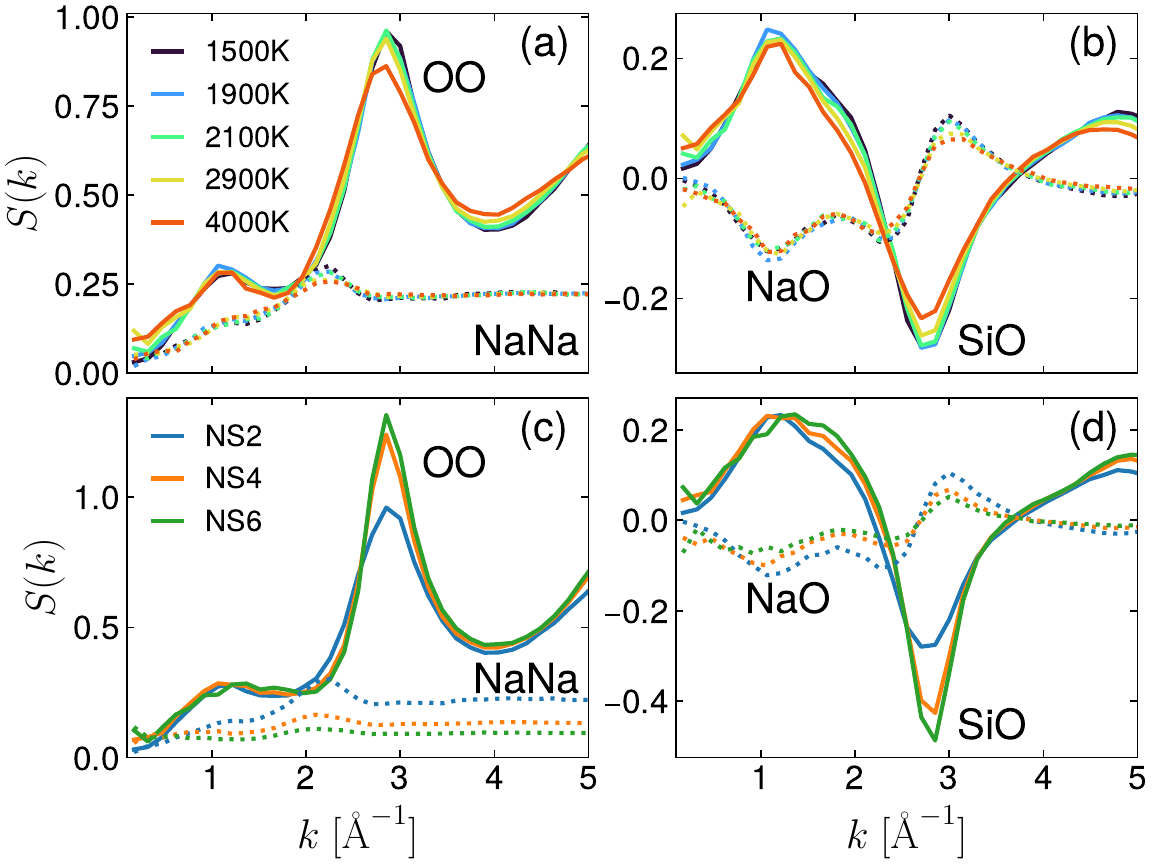}
\caption{Partial structure factors $S_{\alpha\beta}(k)$ of NS2.
Panels (a) and (b) show the temperature dependence of $S_\mathrm{OO}(k)$ and $S_\mathrm{NaNa}(k)$ (panel (a)) and $S_\mathrm{SiO}(k)$ and $S_\mathrm{NaO}(k)$ (panel (b)).
Panels (c) and (d) show the corresponding compositional dependence at \qty{1500}{\kelvin}.}
\label{fig:sk}
\end{figure}

We begin our structural analysis by examining the partial structure factors $S_{\alpha\beta}(k)$ between $\alpha$ and $\beta$ atoms, which provide a quantitative description of the spatial correlations in the sodium silicate melt.
The structure factor $S_{\alpha\beta}(k)$ is defined as~\cite{HansenMcDonald}
\begin{align}
S_{\alpha \beta}(k) = \frac{1}{\sqrt{N_\alpha N_\beta}} \Braket{\rho_\alpha\pab{\bm{k}} \rho_\beta\pab{-\bm{k}}},
\end{align}
where $\rho_\alpha\pab{\bm{k}} = \sum_{j=1}^{N_\alpha} \exp \pab{i \bm{k} \cdot \bm{r}_j}$ is the Fourier transform of the microscopic density at wave vector $\bm{k}$.
We use the \texttt{freud} package~\cite{freud2020} for this calculation.
Figure~\ref{fig:sk} illustrates the evolution of these correlations as a function of temperature and composition.

The temperature dependence of the partial structure factors, presented in Figs.~\ref{fig:sk} (a) and (b) for the NS2 composite, reveals a systematic intensification and sharpening of principal peaks as the system is cooled from \qty{4000}{K} to \qty{1500}{K}.
This trend signifies a progressive enhancement of structural order and a reduction in thermal fluctuations, allowing the underlying glassy framework to emerge with increasing definition.
Notably, the peaks associated with the network-forming species, specifically the $S_\mathrm{OO}(k)$ maximum and the pronounced $S_\mathrm{SiO}(k)$ minimum, exhibit a robust development, suggesting that the silicate framework establishes its fundamental topological features well above the glass transition.
In contrast, the correlations involving \ce{Na} ions, $S_\mathrm{NaNa}(k)$ and $S_\mathrm{NaO}(k)$, whilst also sharpening upon cooling, remain relatively broader than those of the \ce{SiO2} framework even at the lowest temperature.
This persistence of the broad distribution of $S(k)$ indicates that the alkali modifiers retain a significant degree of local structural flexibility and disordered liquid-like structure within the interstitial voids, even as the surrounding \ce{SiO2} network proceeds to structural freezing.

The compositional influence on the structural arrangement, shown in Figs.~\ref{fig:sk} (c) and (d), highlights the dual role of \ce{Na} as both a network modifier and a self-organising species.
A particularly striking feature is the emergence and systematic growth of a prepeak in $S_\mathrm{NaNa}(k)$ at $k \approx \qty{1}{\per\angstrom}$ as the sodium fraction increases from NS6 to NS2.
This low-$k$ feature constitutes direct evidence for intermediate-range order, demonstrating that the \ce{Na} ions do not disperse randomly but instead organise into spatially correlated regions or channels within the silica matrix~\cite{Meyer2004Channel}.
Simultaneously, the intensification of the \ce{Na}--\ce{Na} prepeak is accompanied by a discernible damping of the $S_\mathrm{OO}(k)$ and $S_\mathrm{SiO}(k)$ correlations.
This decrease reflects the systematic de-polymerisation of the silicate framework;
as the alkali concentration rises, the \ce{Na} ions act to sever the \ce{Si}--\ce{O} linkages, thereby reducing the overall connectivity of the network.
This structural degradation is further corroborated by the enhancement of the $S_\mathrm{NaO}(k)$ correlation, which indicates an increased population of \ce{Na} ions coordinating with oxygens that are not bridging \ce{Si}.
Taken together, these observations suggest a spatial segregation wherein the alkali ions actively carve out their own preferential pathways by locally disrupting the covalent network, a structural precursor to the kinetic decoupling discussed in Secs.~\ref{sec:relaxation} and \ref{sec:DH}.

\subsection{Bond-bridging oxygens}
\begin{figure}
\centering
\includegraphics[width=\linewidth]{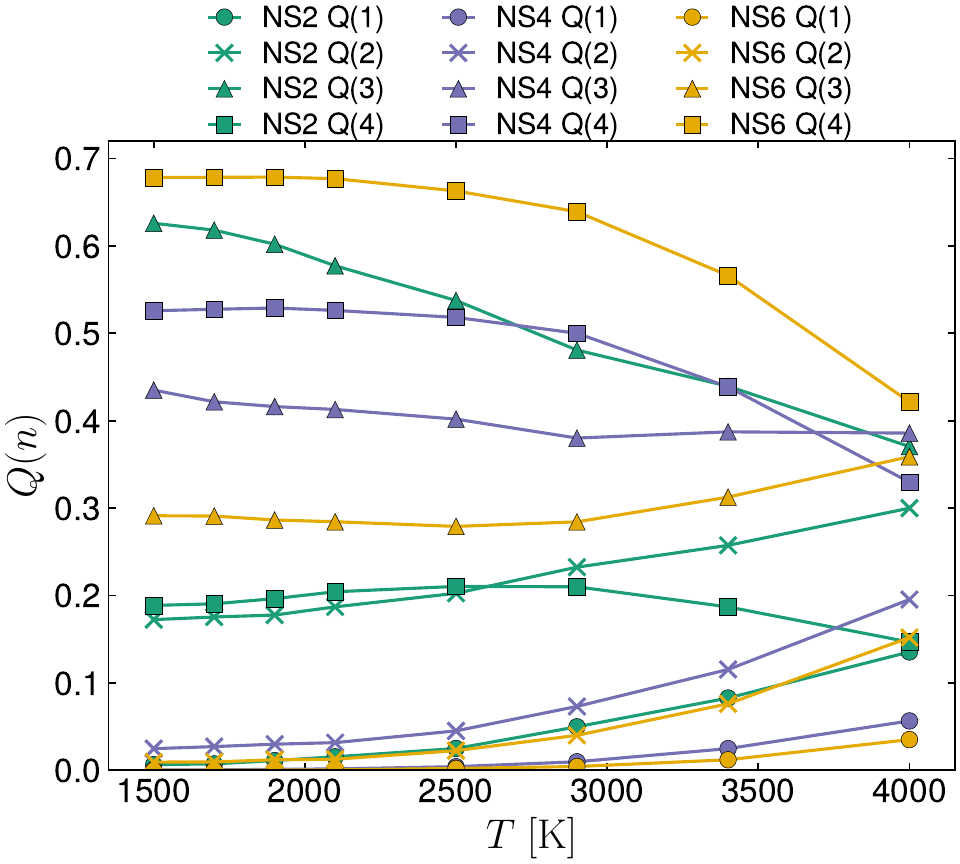}
\caption{Bond-bridging oxygen $Q(n)$ of every composites.}
\label{fig:bonding_oxygen}
\end{figure}

To further elucidate the topological nature of the silicate framework, we now examine the distribution of $Q(n)$ species.
$Q(n)$ is defined as the fraction of \ce{SiO4} tetrahedra possessing $n$ bridging oxygens and is regularly measured in both experimental~\cite{STEBBINS1988359,Farnan1990,MAEKAWA199153} and numerical studies~\cite{HORBACH200187}.
Figure~\ref{fig:bonding_oxygen} illustrates a compositional trend in this distribution across the investigated temperatures.
As the sodium concentration increases from NS6 to NS2, there is a pronounced shift from the predominantly fully-connected $Q(4)$ environment, typical of pure silica, towards fragmented $Q(3)$ and $Q(2)$ species.
This distribution shift provides quantitative evidence for the systematic de-polymerisation of the silica framework driven by the alkali ions.
This fragmented structure, particularly prominent in NS2 with high $Q(3)$ and $Q(2)$ populations, results in a more open network with increased interstitial volume.
These structural features represent the local realisation of the preferential pathways or channels implied by the $S_\mathrm{NaNa}(k)$ prepeak discussed in the previous section (Fig.~\ref{fig:sk} (c)).
Such a framework serves as a direct structural precursor for the accelerated bulk relaxation observed in sodium-rich composites, and critically for the remarkable decoupling of sodium ions from the slower silica matrix.
Moreover, the co-existence of disparate local $Q(n)$ environments provides a strong structural basis for the high degree of dynamic heterogeneity observed for framework atoms, with a non-uniform spatial distribution of local constraints.

Regarding the temperature dependence, while rapid variations in $Q(n)$ populations are evident in the high-temperature regime, there is a clear trend towards a suppression of the variation as the temperature decreases from \qty{4000}{K} to \qty{1500}{K}.
This behaviour signifies the suppression of dynamic bond rearrangements and the progressive fixation of a topological network, typical of the glass formation process.
Although this is natural in the glass formation process upon cooling, there are several differences between the previous report with empirical potential~\cite{HORBACH200187}.
While the previous data shows that $Q(2)$ or $Q(4)$ vary their values as lowering temperature at least until $T = \qty{2100}{\kelvin}$ for NS2, our $Q(2)$ or $Q(4)$ of NS2 reach their plateaus at around $T \approx \qty{2500}{\kelvin}$.
We attribute this discrepancy in part to incomplete equilibration at lower temperatures in the present simulations, a limitation that warrants further investigation.

\clearpage
\onecolumngrid

\setcounter{section}{0}
\setcounter{figure}{0}
\setcounter{table}{0}
\setcounter{equation}{0}

\renewcommand{\thefigure}{S\arabic{figure}}
\renewcommand{\thetable}{S\arabic{table}}
\renewcommand{\theequation}{S\arabic{equation}}

\section*{Supplementary Information}
\subsection*{Pair correlation function}
\begin{figure*}[b]
\centering
\includegraphics[width=\linewidth]{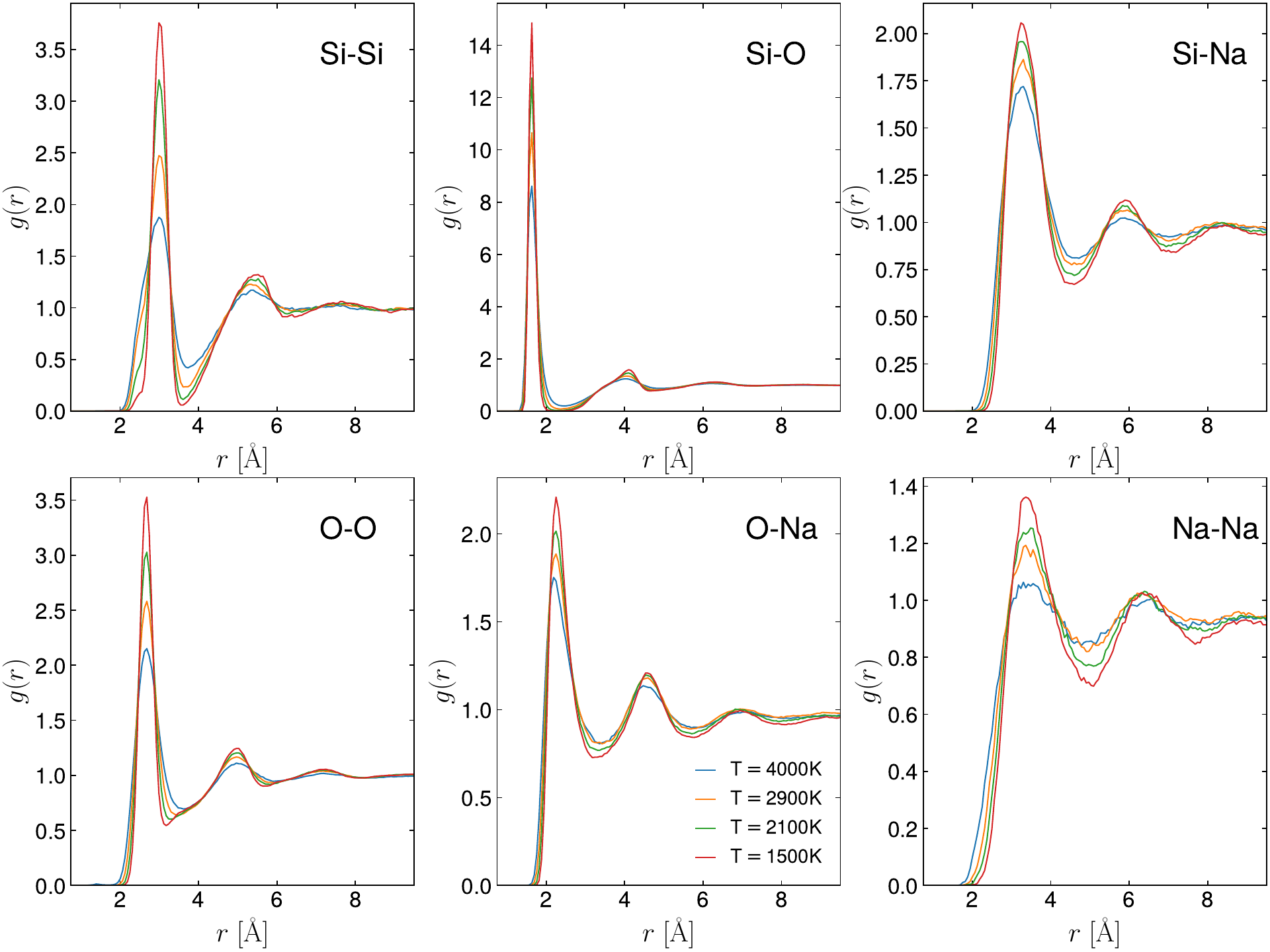}
\caption{Pair correlation functions $g_{\alpha\beta}(r)$ for NS2.}
\label{fig:rdfNS2}
\end{figure*}

\begin{figure*}
\centering
\includegraphics[width=\linewidth]{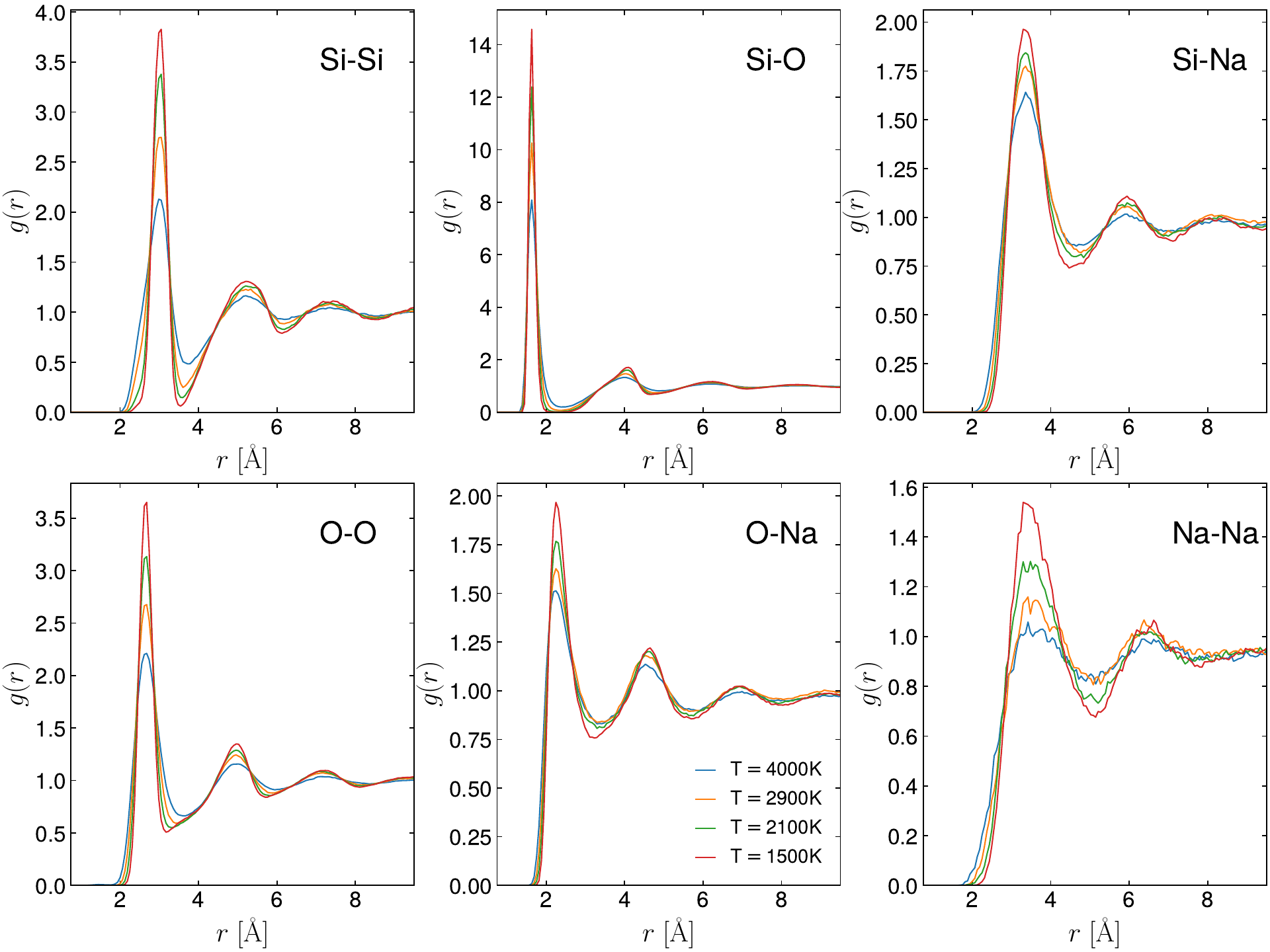}
\caption{Pair correlation functions $g_{\alpha\beta}(r)$ for NS4.}
\label{fig:rdfNS4}
\end{figure*}

\begin{figure*}
\centering
\includegraphics[width=\linewidth]{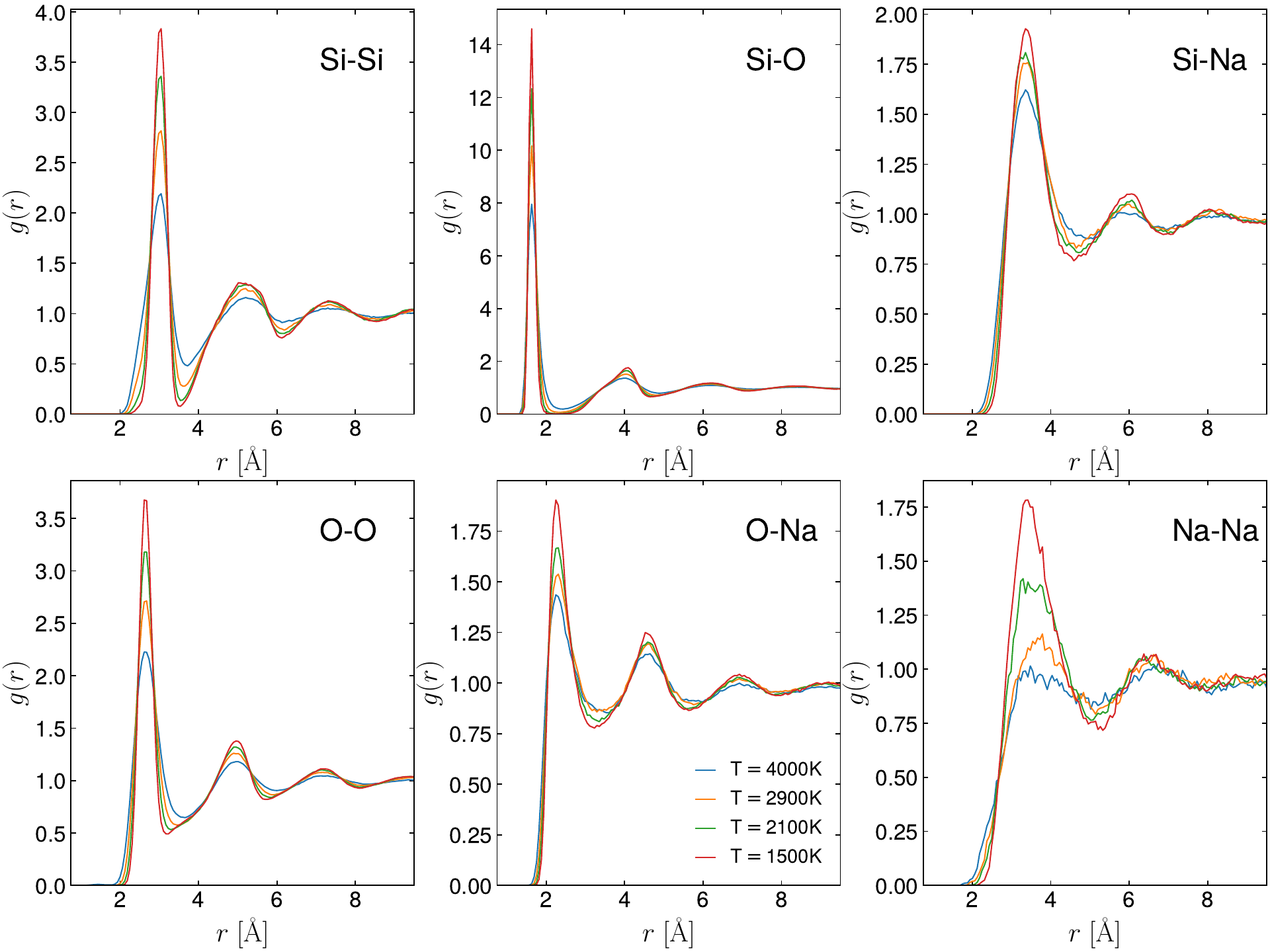}
\caption{Pair correlation functions $g_{\alpha\beta}(r)$ for NS6.}
\label{fig:rdfNS6}
\end{figure*}

In this Supplementary Information, we present the pair correlation function $g_{\alpha\beta}(r)$ (with $\alpha, \beta = \ce{Si}, \ce{O}, \ce{Na}$)
\begin{align}
g_{\alpha\beta}(r) = \frac{V}{N_\alpha\pab{N_\beta - \delta_{\alpha\beta}}}\Braket{\sum_{i=1}^{N_\alpha} \sum_{j=1}^{N_\beta} \frac{1}{4\pi r^2} \delta \pab{r - \lvert \bm{r}_i - \bm{r}_j \rvert}},
\end{align}
where $V$ is the total volume of the system~\cite{HansenMcDonald}.
We use the \texttt{freud} package~\cite{freud2020} for the calculation of $g_{\alpha\beta}(r)$.
Figures~\ref{fig:rdfNS2}--\ref{fig:rdfNS6} show the all cases of $g_{\alpha\beta}(r)$ for NS2, NS4, and NS6, respectively.

\end{document}